\documentclass[a4paper,12pt]{article}
\usepackage{amsmath}
\usepackage{amssymb}
\usepackage[dvipdfmx]{graphicx}
\usepackage{bm}

\usepackage{wrapfig}
\usepackage[nosort]{cite}

\newcommand{\nn}{\nonumber}
\newcommand{\tr}{\text{tr}\,}

\newcommand{\vev}[1]{\left\langle #1 \right\rangle}
\newcommand{\dvev}[1]{\left\langle\!\left\langle #1 \right\rangle\!\right\rangle}

\newcommand{\be}{\begin{equation}}
\newcommand{\ee}{\end{equation}}
\newcommand{\bea}{\begin{eqnarray}}
\newcommand{\eea}{\end{eqnarray}}
\newcommand{\cO}{{\cal O}}
\newcommand{\binomi}[2]{\begin{pmatrix} #1 \\ #2 \end{pmatrix}}
 \def\XXint#1#2#3{{\setbox0=\hbox{$#1{#2#3}{\int}$}
 \vcenter{\hbox{$#2#3$}}\kern-.5\wd0}}

\begin{document}
\thispagestyle{empty} \addtocounter{page}{-1}
\vspace*{1cm}

\begin{center}
{\large \bf Resurgence of one-point functions  \\
in a matrix model for 2D type IIA superstrings}\\
\vspace*{2cm}
Tsunehide Kuroki$^{*,\ddagger}$ and Fumihiko Sugino$^\dagger$\\
\vskip0.7cm
{}$^*${\it General Eduction, National Institute of Technology, Kagawa College}\\
\vspace*{1mm}
{\it 551 Kohda, Takuma-cho, Mitoyo, Kagawa 769-1192, Japan}\\
\vspace{1mm}
{}$^\ddagger${\it Osaka City University Advanced Mathematical Institute (OCAMI)} \\
\vspace*{1mm}
{\it 3-3-138 Sugimoto, Sumiyoshi-ku, Osaka 558-8585, Japan}\\
\vspace*{0.2cm}
{\tt t96ki@hotmail.com}\\
\vskip0.4cm
{}$^\dagger${\it Center for Theoretical Physics of the Universe, Institute for Basic Science (IBS)} \\
\vspace*{1mm}
{\it Expo-ro 55, Yuseong-gu, Daejeon 34126, Republic of Korea}\\
\vspace*{0.2cm}
{\tt fusugino@gmail.com}\\
\end{center}
\vskip1.5cm
\centerline{\bf Abstract}
\vspace*{0.3cm}
{\small 
In the previous papers, the authors pointed out correspondence between 
a supersymmetric double-well matrix model and two-dimensional type IIA superstring theory 
on a Ramond-Ramond background. 
This was confirmed by agreement between planar correlation functions in the matrix model 
and tree-level amplitudes in the superstring theory. 
Furthermore, in the matrix model we computed one-point functions of single-trace operators 
to all orders of genus expansion in its double scaling limit, and found that the large-order behavior of 
this expansion is stringy and not Borel summable. 
In this paper, we discuss resurgence structure of these one-point functions 
and see cancellations of ambiguities in their trans-series. 
More precisely, we compute both series of ambiguities arising in a zero-instanton sector 
and in a one-instanton sector, and confirm how they cancel each other. 
In case that the original integration contour is a finite interval not passing through 
a saddle point, we have to choose an appropriate integration path in order for resurgence to work. 
}
\vspace*{1.1cm}

\newpage

%\maketitle  IS IGNORED %%%%%%%%%%%
%%%%%%%%%%%%%%%%%%%%%%%%%%%%%%%%%%%%%%%%%%%%%%%%%%%%%%%%%%%%%%%%%%%%%%%%%%%%%%
\section{Introduction}
\label{sec:intro}
\setcounter{equation}{0}
%%%%%%%%%%%%%%%%%%%%%%%%%%%%%%%%%%%%%%%%%%%%%%%%%%%%%%%%%%%%%%%%%%%%%%%%%%%%%%
Resurgence \cite{Ec1,Pham1,BH1,Howls1,DH1,Costin1,Sauzin1,Sauzin2} 
has attracted lots of attention by its intriguing property to make intimate connection 
between perturbative and nonperturbative quantities. 
From data of higher-order perturbative expansion, resurgence enables us to extract nonperturbative aspects. 
It has been investigated how resurgence works in each of various quantum theories\footnote
{Resurgence structure has been studied in various models and theories based on several motivations: 
see e.g. in quantum mechanics 
\cite{Alvarez1,ZinnJustin:2004ib,Dunne:2013ada,
Escobar-Ruiz:2015nsa,Misumi:2015dua,Behtash:2015zha,Gahramanov:2015yxk,
Fujimori:2016ljw,Fujimori:2017oab,Dunne:2016jsr,Serone:2016qog,Basar:2017hpr,Alvarez:2017sza}, 
string theories \cite{Marino:2008vx,Marino:2006hs,Grassi:2014cla} 
as well as quantum field theories \cite{Marino:2012zq,Aniceto:2015rua,
Dunne:2012ae,Cherman:2013yfa,Misumi:2014jua,Nitta:2014vpa,Behtash:2015kna,
Dunne:2015ywa,Buividovich:2015oju,Demulder:2016mja,Sulejmanpasic:2016llc,
Gukov:2016njj,Gang:2017hbs,Argyres:2012vv,Dunne:2015eoa,Yamazaki:2017ulc,
Russo:2012kj,Aniceto:2014hoa,Honda:2016mvg,Dorigoni:2017smz,Honda:2017cnz,
Fujimori:2018nvz}.
}, 
whereas we still do not know much about a unified picture or classification 
of their resurgence structure\footnote
{Recent progress in this direction have been made, e.g. in \cite{Basar:2017hpr}}. 
In addition, even in a specific model or theory, we have not clarified how resurgence structure 
changes 
depending on its physical observables. Namely, some observables may be strongly affected 
by nonperturbative effects and resurgence plays an important role in extracting nonperturbative information, 
while other observables may not and their perturbative series may behave too well to get insight 
into nonperturbative aspects by resurgent analysis. 

Furthermore, we have not known much about relation between resurgence structure and physics. 
In general, nonperturbative dynamics is important to explain how 
interesting physical phenomena like quark confinement or symmetry breaking take place. 
When resurgence extracts some 
information on nonperturbative effects from perturbation theory, 
we expect that some insights into such interesting physics are available. 
Thus, it is desirable to find various examples where resurgence helps us understand 
nonperturbative aspects of physics. 

Based on these motivations, in this paper we study resurgence structure 
in a supersymmetric double-well matrix model with the action 
\begin{align}
S = N \tr \left[\frac12 B^2 +iB(\phi^2-\mu^2) +\bar\psi (\phi\psi+\psi\phi)\right], 
\label{eq:S}
\end{align}
where $B$ and $\phi$ are $N\times N$ Hermitian matrices, 
and $\psi$ and $\bar\psi$ are $N\times N$ Grassmann-odd matrices. 
$\mu^2$ is a parameter of the model. 
The action $S$ is invariant under supersymmetry transformations generated by $Q$ and $\bar{Q}$: 
\begin{align}
&Q\phi =\psi, &&Q\psi=0, &&Q\bar{\psi} =-iB, &&QB=0, \nn \\
&\bar{Q} \phi = -\bar{\psi}, &&\bar{Q}\bar{\psi} = 0, 
&&\bar{Q} \psi = -iB, &&\bar{Q} B = 0,  
\label{eq:SUSY}
\end{align}
which lead to the nilpotency: $Q^2=\bar{Q}^2=\{ Q,\bar{Q}\}=0$. 
One of the most interesting features of this model 
is that the supersymmetry is spontaneously broken by nonperturbative effects 
in a certain large-$N$ limit called as double scaling limit 
(defined in \eqref{eq:dsl})~\cite{Endres:2013sda,Nishigaki:2014ija}. We also have proposed 
in \cite{Kuroki:2013qpa} that this model under the double scaling limit gives nonperturbative 
formulation of type IIA superstring theory in two dimensions on a Ramond-Ramond background. 
From these, we can regard this model as an invaluable example of spontaneously broken target-space supersymmetry in string theory. 
In this paper, we concentrate on one-point functions of powers of matrix $\phi$: 
$\vev{\frac1N\tr\phi^n}$. In the previous work \cite{Kuroki:2012nt}, it is shown that 
the operators with $n$ even ($n\in 2\bm N$) are essentially supersymmetric, 
and $1/N$ or genus expansions of their one-point functions are polynomials in the parameter $\mu^2$, 
terminating at some genus, which do not lead to any nonanalytic behavior in the double scaling limit. 
In \cite{Endres:2013sda}, taking account of effects nonperturbative in $1/N$, 
we have calculated the one-point function $\vev{\frac1N\tr (\phi^2-\mu^2)}$ 
(or equivalently $\vev{\frac{1}{N}\tr B}$) 
as an order parameter of the spontaneous supersymmetry breaking. On the other hand, 
the odd-power operators ($n\in 2\bm N-1$) are not supersymmetric, and genus expansions 
of their one-point functions exhibit stringy growth of the expansion coefficients as $(2h)!$ as genus $h$ grows. 
In this paper, we consider the one-point functions of the odd-power operators and study 
their resurgence structure. 
Since instantons in the matrix model \eqref{eq:S} contribute to the order parameter $\vev{\frac1N\tr (\phi^2-\mu^2)}$ 
and trigger spontaneous supersymmetry breaking~\cite{Endres:2013sda},
our main interest is to clarify 
how resurgence structure for the odd-power operators is related to such nonperturbative physics. 

Another advantage of considering our model \eqref{eq:S} is that the existence of 
the Nicolai mapping~\cite{Endres:2013sda}. 
Although resurgence requires data of large-order perturbation series, it is not easy to obtain such data in general. 
In case that a theory is supersymmetric, 
we may compute its perturbative expansion to all orders, 
but in turn it may be Borel summable and have trivial resurgence structure. 
One of nice approaches to overcome this issue is to introduce a parameter explicitly breaking 
supersymmetry~\cite{Fujimori:2016ljw,Fujimori:2017oab,Dunne:2016jsr}. 
In \cite{Kuroki:2016ucm} and this paper, 
we propose another way to obtain perturbative expansion to all orders for resurgence: 
we consider nonsupersymmetric quantities in a supersymmetric model. 
In fact, even in calculation of nonsupersymmetric quantities in \eqref{eq:S}, 
the Nicolai mapping is available reflecting the existence of supersymmetry in the action. 
This kind of idea will be useful, in particular, in supersymmetric field theories 
as in \cite{Russo:2012kj,Aniceto:2014hoa,Honda:2016mvg,Dorigoni:2017smz,Honda:2017cnz}. 
 
The organization of this paper is as follows. In the next section, we give a brief review 
of the supersymmetric double-well matrix model. Correlation functions are expressed 
in terms of eigenvalues of the matrix $\phi$ and are defined in each instanton sector. 
In section \ref{sec:Nicolai}, we explain how to compute the one-point functions 
of odd powers of $\phi$ by utilizing the Nicolai mapping. 
In section \ref{sec:zero-inst}, we consider contribution from the zero-instanton sector 
to the one-point functions, and find that there exists a series of ambiguities 
after applying the Borel resummation 
technique. Then, in section \ref{sec:one-inst} we see that contribution from the one-instanton sector 
also has another series of ambiguities, and confirm that these series exactly cancel each other 
at the leading and next-to-leading orders. 
The last section is devoted to conclusions and discussions.

%%%%%%%%%%%%%%%%%%%%%%%%%%%%%%%%%%%%%%%%%%%%%%%%%%%%%%%%%%%%%%%%%%%%%%%%%%%%%
\section{Review of the supersymmetric matrix model}
\label{sec:MM_review}
\setcounter{equation}{0}
%%%%%%%%%%%%%%%%%%%%%%%%%%%%%%%%%%%%%%%%%%%%%%%%%%%%%%%%%%%%%%%%%%%%%%%%%%%%%
In this section, we give a brief review of the supersymmetric double-well matrix model defined by the action (\ref{eq:S}), 
which has been proposed as a nonperturbative formulation of type IIA superstring theory 
in two dimensions. 

%%%%%%%%%%%%%%%%%%%%%%%%%%%%%%%%%%%%%%%%%
\subsection{Supersymmetry and large-$N$ limit}
After integrating out the auxiliary variable $B$ in (\ref{eq:S}), 
the scalar potential of $\phi$ reads
\begin{align}
V(\phi)=\frac12(\phi^2-\mu^2)^2. 
\label{eq:DW}
\end{align}
In the planar limit ($N\rightarrow\infty$ with $\mu^2$ fixed) 
of the matrix model, there are infinitely degenerate supersymmetric vacua 
parametrized by filling fractions $(\nu_+, \nu_-)$ for $\mu^2\geq 2$. 
The filling fractions represent configurations that $\nu_\pm N$ of the eigenvalues of $\phi$ are around the minimum 
$\pm |\mu |$ of the double-well potential \eqref{eq:DW} \cite{Kuroki:2009yg, Kuroki:2010au}. 
On the other hand, for $\mu^2<2$ we have a unique vacuum which breaks the supersymmetry. 
The boundary $\mu^2=2$ is a critical point at which the third-order phase transition occurs. 
In the planar limit, it is explicitly seen in~\cite{Kuroki:2013qpa,Kuroki:2012nt} that 
the result of several types of correlation functions in the matrix model reproduces 
the tree amplitudes in two-dimensional type IIA superstring theory 
on a nontrivial Ramond-Ramond background.  
In addition, we have considered the following double scaling limit~\cite{Endres:2013sda} that 
approaches the critical point from the inside of the supersymmetric phase: 
\begin{align}
N\rightarrow\infty, \quad \mu^2\rightarrow 2+0, \quad  \mbox{with} \quad  
s=N^{\frac23}(\mu^2-2):\mbox{~fixed}.
\label{eq:dsl}
\end{align}
This limit of the matrix model is expected to provide a nonperturbative formulation 
of the superstring theory with string coupling constant $g_s$ proportional to $s^{-\frac32}$. 
{}From this viewpoint the planar limit mentioned above is regarded as $g_s\rightarrow 0$ limit. 
In fact, in \cite{Kuroki:2016ucm} one-point functions for the single-trace operators of powers 
of $\phi$ are explicitly calculated at arbitrary genus and found to be finite at each genus 
under the double scaling limit \eqref{eq:dsl}. 
In~\cite{Endres:2013sda,Nishigaki:2014ija}, contribution 
from matrix-model instantons (isolated eigenvalues of $\phi$ located 
around the top of the effective potential) to the free energy is found to be also finite 
and to have a factor $\exp\left(-C/g_s\right)$ 
with a constant $C$ of $\cO(1)$. 
This form is typical of solitonic objects in string theory (D-branes). 
The correspondence between isolated eigenvalues and solitons is also observed 
in well-established bosonic noncritical string theories 
\cite{David:1992za,Kazakov:2004du,Hanada:2004im,Kawai:2004pj,Sato:2004tz,Ishibashi:2005dh,Ishibashi:2005zf,
Kuroki:2006wn}. 
The most interesting feature of the model ever found is that these instantons cause 
spontaneous supersymmetry breaking in the matrix model, 
which implies violation of target-space supersymmetry 
induced by nonperturbative effects in the corresponding superstring theory.   
The aim of this paper is to investigate connection between contribution from higher genus  
in the absence of the instanton (in the zero-instanton sector) and 
that from the one-instanton sector through the one-point functions. 

%%%%%%%%%%%%%%%%%%%%%%%%%%%%%%%%%%%%%
\subsection{Correlation functions in fixed filling fraction}
In this subsection, we define correlation functions of our model \eqref{eq:S} 
in a fixed filling fraction. 
First, the partition function is expressed in terms of the eigenvalues of $\phi$ as follows 
\cite{Endres:2013sda,Kuroki:2012nt}: 
\bea
Z& \equiv&(-1)^{N^2}\int d^{N^2}B\,d^{N^2}\phi\,\left(d^{N^2}\psi\,d^{N^2}\bar{\psi}\right)\,e^{-S} \nn \\
&=&\tilde C_N\int_{-\infty}^{\infty}
\left(\prod_{i=1}^N2\lambda_id\lambda_i\right)\triangle(\lambda^2)^2\,
e^{-N\sum_{i=1}^N\frac12(\lambda_i^2-\mu^2)^2}, 
\label{eq:Z}
\eea
where the normalization of the integration measure is fixed as 
\bea
& & \int d^{N^2}\phi\,e^{-N\tr\left(\frac12\phi^2\right)}= \int d^{N^2}B\,e^{-N\tr\left(\frac12 B^2\right)}=1 , \nn \\
& & (-1)^{N^2}\int \left(d^{N^2}\psi\,d^{N^2}\bar{\psi}\right)\,e^{-N\tr \left(\bar{\psi}\psi\right)}=1.
\eea
$\tilde C_N$ is a constant dependent only on $N$: 
$\tilde{C}_N=(2\pi)^{-\frac{N}{2}}N^{\frac{N^2}{2}}\left(\prod_{k=0}^Nk!\right)^{-1}$~\cite{Kuroki:2010au}, and 
$\triangle(x)$ stands for the Vandermonde determinant for eigenvalues $x_i$ ($i=1,\cdots, N$): $\triangle(x) \equiv \prod_{i>j}(x_i-x_j)$. 
By dividing the integration region of each $\lambda_i$ according to the filling fraction, 
the total partition function can be expressed as a sum of 
each partition function with a fixed filling fraction: 
\begin{align}
&Z=\sum_{\nu_-N=0}^{N}\frac{N!}{(\nu_+N)!(\nu_-N)!}\,Z_{(\nu_+,\nu_-)}, \nn \\ 
&Z_{(\nu_+,\nu_-)}\equiv \tilde C_N
\int_0^{\infty}\left(\prod_{i=1}^{\nu_+N}2\lambda_id\lambda_i\right)
\int_{-\infty}^0\left(\prod_{j=\nu_+N+1}^N2\lambda_jd\lambda_j\right)
\triangle(\lambda^2)^2 
\,e^{-N\sum_{m=1}^N\frac12(\lambda_m^2-\mu^2)^2}. 
\label{eq:ZinFFsector}
\end{align}
By changing the integration variables $\lambda_j\rightarrow -\lambda_j$ ($j=\nu_+N+1,\cdots, N$), 
it is easy to find that 
$Z_{(\nu_+, \nu_-)}=(-1)^{\nu_-N}Z_{(1,0)}$
and the total partition function vanishes. 

Next, we define the correlation function of $K$ single-trace operators $\frac{1}{N}\tr \cO_a(\phi)$ 
($a=1,\cdots, K$) in the filling fraction $(\nu_+,\nu_-)$ as 
\begin{align}
\vev{\prod_{a=1}^K\frac{1}{N}\tr{\cal O}_a(\phi)}^{(\nu_+,\nu_-)} &\equiv \frac{\tilde C_N}{Z_{(\nu_+,\nu_-)}}
\int_0^{\infty}\left(\prod_{i=1}^{\nu_+N}2\lambda_id\lambda_i\right)
\int_{-\infty}^0\left(\prod_{j=\nu_+N+1}^N2\lambda_jd\lambda_j\right)\triangle(\lambda^2)^2\nn \\
& \hspace{20mm}\times \left(\prod_{a=1}^K\frac{1}{N}\sum_{i=1}^N{\cal O}_a(\lambda_i)\right)\,
e^{-N\sum_{m=1}^N\frac12(\lambda_m^2-\mu^2)^2},  
\label{eq:correlator}
\end{align}
and express its connected part in the $1/N$-expansion:  
\begin{align}
\vev{\prod_{a=1}^K \frac{1}{N}\tr{\cal O}_a(\phi)}_C^{(\nu_+,\nu_-)}=\sum_{h=0}^\infty \frac{1}{N^{2h+2K-2}}\,
\vev{\prod_{a=1}^K \frac{1}{N}\tr{\cal O}_a(\phi)}_{C,\,h}^{(\nu_+,\nu_-)}.  
\label{eq:vev_MM}
\end{align}
$\vev{\,\cdot\,}_{C,\,h}^{(\nu_+,\nu_-)}$ denotes the connected correlation function on a handle-$h$ random surface 
with the $N$-dependence factored out; i.e., the quantity of ${\cal O}(N^0)$. 
Let us consider the case where $\cO_a(\phi)$ are polynomials of $\phi$. 
Operators $\frac{1}{N}\tr B^k$ or (linear combinations of) 
$\frac{1}{N}\tr\phi^{2k}$ ($k\in\bm N\cup\{0\}$) are invariant 
under the supersymmetries (\ref{eq:SUSY}). 
For these operators, multi-point functions at the planar level ($h=0$) and higher-genus one-point functions 
do not exhibit any nonanalytic behavior as $s\to 0$~\cite{Kuroki:2012nt,Kuroki:2016ucm}, 
which is characteristic of protection by supersymmetry.  
On the other hand, operators of odd powers: $\frac{1}{N}\tr\phi^{2k+1}$ ($k\in\bm N\cup\{0\}$) 
are not invariant under either of $Q$ or $\bar{Q}$, and their correlation functions 
have nontrivial dependence on $s$~\cite{Kuroki:2016ucm} as we will mention in the next section.  
For simplicity, in the following we focus on the one-point function of the odd-power operators: 
\eqref{eq:vev_MM} with $K=1$, ${\cal O}_1(\phi)=\phi^{2k+1}$ 
in the filling fraction\footnote
{It is shown in \cite{Kuroki:2012nt} that at least at the planar level ($h=0$) 
and up to the three-point functions 
($1\leq K\leq 3$), it is easy to recover filling fraction dependence of correlation functions 
from those in $(\nu_+,\nu_-)=(1,0)$.
} $(\nu_+,\nu_-)=(1,0)$. 

%%%%%%%%%%%%%%%%%%%%%%%%%%%%%%%%%%%%%%%%%
\subsection{Correlation functions in fixed instanton sector}
In this subsection, we divide correlation functions in the $(1,0)$ sector into contributions 
from definite instanton numbers as done in~\cite{Endres:2013sda}. 
In \eqref{eq:ZinFFsector}, the partition function $Z_{(1,0)}$ with the filling fraction $(1,0)$ 
is expressed as the integrations of $N$ eigenvalues along the positive real axis. 
The eigenvalue distribution in the planar limit is given as \cite{Kuroki:2012nt,Kuroki:2009yg} 
\begin{align}
\left.\vev{\frac{1}{N}\sum_{i=1}^N\delta(x-\lambda_i)}^{(1,0)}\right|_{\text{planar}}=
\begin{cases}
\frac{x}{\pi}\sqrt{(x^2-a^2)(b^2-x^2)} & (a<x<b) \\
0 & (\text{otherwise}),
\end{cases}
\label{eq:rho}
\end{align}
with $a=\sqrt{\mu^2-2}$ and $b=\sqrt{\mu^2+2}$, which means that 
all the eigenvalues are confined in the interval $[a,b]$. Dividing 
the integration region of each eigenvalue $\bm R_+=[0,\infty)$ into the inside and outside 
of the support:
\begin{align}
\int_0^\infty d\lambda_i=\int_a^b d\lambda_i+\int_{\bm R_+\setminus [a,b]} d\lambda_i, 
\end{align}
we decompose the partition function as 
\begin{align}
&Z_{(1,0)}=\sum_{p=0}^N\left.Z_{(1,0)}\right|_{p\text{-inst.}}, \nn \\
&\left.Z_{(1,0)}\right|_{p\text{-inst.}}={}
\begin{pmatrix}N \\ p\end{pmatrix}\tilde C_N
\int_a^b\prod_{i=1}^{N-p}2\lambda_id\lambda_i
\int_{\bm R_+\setminus [a,b]}\prod_{i=1}^{p}2\lambda_jd\lambda_j\,
\Delta(\lambda^2)^2 
e^{-N\sum_{i=1}^N\frac12(\lambda_i^2-\mu^2)^2}.
\label{eq:instantonsectors}
\end{align}
Each contribution with fixed $p$ is regarded as the partition function in the $p$-instanton sector.  
In fact, an instanton in our model corresponds to a saddle point of effective potential 
$V_{\text{eff}}(\lambda_i)$ with respect to a single eigenvalue $\lambda_i$, 
which is obtained by integrating out all the eigenvalues other than $\lambda_i$ in \eqref{eq:Z}. 
Its saddle point turns out to be the origin $\lambda_i=0$ \cite{Endres:2013sda}. 
For large $s$ (small $g_s$) under the double scaling limit \eqref{eq:dsl}, 
the main contribution from the outside of the support $\bm R_+\setminus [a,b]$ 
is provided by such an instanton located at the origin.  
Then as mentioned in Introduction, leading contribution from the instanton 
takes the form of $\exp\left(-C/g_s\right)$. 
Correlation functions in the $p$-instanton sector can also be defined in a similar manner: 
\begin{align}
\vev{\cO}^{(1,0)}
=\sum_{p=0}^N
\frac{\left.Z_{(1,0)}\right|_{p\text{-inst.}}}{Z_{(1,0)}} 
\left.\dvev{\cO}^{(1,0)}\right|_{p\text{-inst.}}, 
\label{eq:vev_pinst}
\end{align}
where $\left.\dvev{\cO}^{(1,0)}\right|_{p\text{-inst.}}$ denotes an expectation value of $\cO$ within 
the $p$-instanton configurations normalized by $\left.Z_{(1,0)}\right|_{p\text{-inst.}}$.  
According to~\cite{Endres:2013sda,Nishigaki:2014ija}, the partition functions behave as 
\be
\left.Z_{(1,0)}\right|_{\text{0-inst.}}=1,\qquad 
\left.Z_{(1,0)}\right|_{\text{$p$-inst.}}
=\left(\frac{e^{-\frac43s^{\frac32}}}{16\pi s^{\frac32}}\right)^p
\times\left[1+\cO(s^{-\frac32})\right] 
\label{eq:Z_pinst_dsl}
\ee
in the double scaling limit with $s$ finite but large, whereas $\left.\dvev{\cO}^{(1,0)}\right|_{p\text{-inst.}}$ has no such exponential suppression. 
Hence \eqref{eq:vev_pinst} is a trans-series expanded by the instanton weight 
$e^{-\frac43s^{\frac32}}/(16\pi s^{\frac32})$:
\begin{align}
\vev{\cO}^{(1,0)} = & \left.\dvev{\cO}^{(1,0)}\right|_{\text{0-inst.}} \nn \\
&+\left.Z_{(1,0)}\right|_{\text{1-inst.}} 
\left(\left.\dvev{\cO}^{(1,0)}\right|_{\text{1-inst.}}
-\left.\dvev{\cO}^{(1,0)}\right|_{\text{0-inst.}}\right) \nn \\
&+\left.Z_{(1,0)}\right|_{\text{2-inst.}} 
\left(\left.\dvev{\cO}^{(1,0)}\right|_{\text{2-inst.}}
-\left.\dvev{\cO}^{(1,0)}\right|_{\text{0-inst.}}\right) \nn \\
&+\left(\left.Z_{(1,0)}\right|_{\text{1-inst.}}\right)^2 
\left(-\left.\dvev{\cO}^{(1,0)}\right|_{\text{1-inst.}}
+\left.\dvev{\cO}^{(1,0)}\right|_{\text{0-inst.}}\right)  \nn \\
&+(\mbox{contribution from the total instanton number $p\geq 3$}), 
\label{eq:O_exp_inst}
\end{align} 
where the first line on the r.h.s has no exponential suppression, 
while other lines have according to \eqref{eq:Z_pinst_dsl}. 
In the previous work \cite{Kuroki:2016ucm}, we have computed the one-point function $\left.\dvev{\cO}^{(1,0)}\right|_{\text{0-inst.}}$ 
to all orders in expansion by $g_s^2\propto s^{-3}$ for $\cO=\tr\phi^n$. 
In what follows we will find that for odd $n$, the Borel resummation of this expansion 
has a series of ambiguities, 
but explicitly show that it is indeed canceled by another series of ambiguities 
arising in the one-instanton sector given in the second line in \eqref{eq:O_exp_inst} 
up to the next-to-leading order.  
This provides a strong support that there is no ambiguity in the trans-series form 
up to the instanton number one. 

In the following, we use the notation: 
\bea
\left. \vev{\cO}^{(1,0)} \right|_{\text{0-inst.}} &= & \left.\dvev{\cO}^{(1,0)}\right|_{\text{0-inst.}} , \nn \\
\left. \vev{\cO}^{(1,0)} \right|_{\text{1-inst.}} &= & \left.Z_{(1,0)}\right|_{\text{1-inst.}} 
\left(\left.\dvev{\cO}^{(1,0)}\right|_{\text{1-inst.}}
-\left.\dvev{\cO}^{(1,0)}\right|_{\text{0-inst.}}\right).
\eea

%%%%%%%%%%%%%%%%%%%%%%%%%%%%%%%%%%%%%%%%%%%%%%%%%%%%%%%%%%%%%%%%%%%%%%%%%%%%%
\section{One-point functions via Nicolai mapping}
\label{sec:Nicolai}
\begin{figure}[h]
\centering
\includegraphics[width=10cm, bb=0 0 750 750, clip]{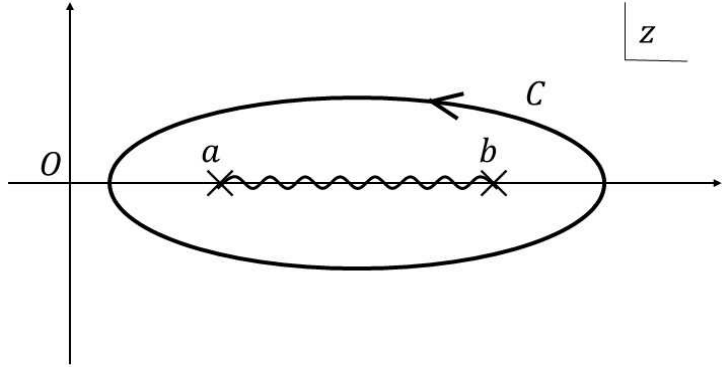}
\vspace{-4cm}
\caption{\small Integration contour $C$ on the complex $z$-plane.}
\label{fig:zcontour}
\end{figure}

\setcounter{equation}{0}
%%%%%%%%%%%%%%%%%%%%%%%%%%%%%%%%%%%%%%%%%%%%%%%%%%%%%%%%%%%%%%%%%%%%%%%%%%%%%
In this section, following the derivation in \cite{Kuroki:2016ucm}, we express the one-point function 
$\vev{\frac{1}{N}\tr\phi^{2k+1}}^{(1,0)}$ in terms of quantities in the Gaussian matrix model. 
Let us first consider the $\phi^2$-resolvent 
\begin{align}
\vev{R_2(z^2)}^{(1,0)}=\vev{\frac1N\tr\frac{1}{z^2-\phi^2}}^{(1,0)}. 
\label{eq:R2}
\end{align}
In terms of the eigenvalues, $R_2(z^2)$ becomes  
\begin{align}
\frac1N\sum_{i=1}^N \frac{1}{z^2-\lambda_i^2}
= \frac1N\frac{1}{2z}\sum_{i=1}^N\left(\frac{1}{z-\lambda_i}+\frac{1}{z+\lambda_i}\right), 
\end{align}
and $1/\left(z-\lambda_i\right)$ ($1/\left(z+\lambda_i\right)$) has poles only 
on the positive (negative) real axis for the filling fraction $(1,0)$. 
This leads to 
\begin{align}
\vev{\frac{1}{N}\tr\phi^{2k+1}}^{(1,0)}
= \oint_{C_0} \frac{dz}{2\pi i}\,2z^{2k+2}\vev{R_2(z^2)}^{(1,0)},
\label{eq:1ptfull}
\end{align}
where $C_0$ is a contour which encloses only the poles at $z=\lambda_i$ for ${}^\forall i$ 
counterclockwise. 
In particular, in the zero-instanton sector, $\lambda_i$'s are all confined in the interval $[a,b]$, 
and therefore, 
\begin{align}
\left.\vev{\frac{1}{N}\tr\phi^{2k+1}}^{(1,0)}\right|_{0-\text{inst.}}
= \oint_C \frac{dz}{2\pi i}\,2z^{2k+2}\vev{R_2(z^2)}^{(1,0)},   
\label{eq:1ptin0inst}
\end{align}
where $C$ denotes a contour encircling the interval $[a,b]$ counterclockwise 
as depicted in Fig.~\ref{fig:zcontour}. 
$C$ does not contain 
$z=0$ inside, and hence contribution from the instanton at the origin is not included 
in \eqref{eq:1ptin0inst}. 
%%%%%%%%%%%%%%%%%%%%%%%%%%%%%%%%%%%%%%%%%%%%%%%%%%%%%%%%%%%%%%%%%%%%%%%
%%%%%%%%%%%%%%%%%%%%%%%%%%%%%%%%%%%%%%%%%%%%%%%%%%%%%%%%%%%%%%%%%%%%%%%%
Note that the $\phi^2$-resolvent is mapped to the resolvent in the Gaussian matrix model. 
In fact, the Nicolai mapping $x_i=\mu^2-\lambda_i^2$ ($i=1,\cdots, N$) recasts 
the partition function $Z_{(1,0)}$ and the one-point function (\ref{eq:correlator}) with $K=1$ 
in the filling fraction $(1,0)$ as 
\begin{align}
Z_{(1,0)}= \tilde{C}_N\int_{-\infty}^{\mu^2}\left(\prod_{i=1}^Ndx_i\right) \triangle(x)^2\,e^{-N\sum_{i=1}^N\frac12x_i^2} \,\equiv Z^{(\text{G'})} 
\label{eq:Z(1,0)_Nicolai} 
\end{align}
and 
\begin{align}
\vev{\frac{1}{N}\tr\phi^{2k+1}}^{(1,0)} =& \frac{\tilde C_N}{Z^{(\text{G'})}}
\int_{-\infty}^{\mu^2}\left(\prod_{i=1}^Ndx_i\right)\triangle(x)^2
\left(\frac{1}{N}\sum_{i=1}^N\left(\mu^2-x_i\right)^{k+\frac12}\right)
e^{-N\sum_{i=1}^N\frac12x_i^2},  
\label{eq:correlator_Nicolai}
\end{align}
respectively. 
Differently from the standard Gaussian matrix model, 
the integrals of the eigenvalues $x_i$ are not over the entire real axis, 
but are bounded from the above by $\mu^2$. 
The superscript $(\text{G'})$ indicates a quantity in such a Gaussian matrix model. 
By introducing an  $N\times N$ Hermitian matrix $M$ whose eigenvalues are $x_i$ ($i=1,\cdots,N$), 
(\ref{eq:correlator_Nicolai}) can be written as 
\begin{align}
\vev{\frac{1}{N}\tr\phi^{2k+1}}^{(1,0)}
=-\oint_{C_0} \frac{dz}{2\pi i}\,2z^{2k+2}\vev{R_M(\mu^2-z^2)}^{(\text{G'})}, 
\label{eq:1ptfullbyGMM} 
\end{align}
and 
\begin{align}
\left.\vev{\frac{1}{N}\tr\phi^{2k+1}}^{(1,0)}\right|_{0-\text{inst.}}
=-\oint_C \frac{dz}{2\pi i}\,2z^{2k+2}\vev{R_M(\mu^2-z^2)}^{(\text{G'})}, 
\label{eq:1ptin0instbyGMM} 
\end{align}
where $R_M(x)\equiv \frac{1}{N}\tr\frac{1}{x-M}$ and the expectation value $\vev{\cdot}^{(\text{G'})}$ is taken in the 
Gaussian matrix model (\ref{eq:Z(1,0)_Nicolai}). 

It is also useful to express the one-point function by introducing the orthogonal polynomials 
$P_n(x)$ ($n=0,1,\cdots$) associated with the Gaussian matrix model 
\eqref{eq:Z(1,0)_Nicolai} \cite{Endres:2013sda}: 
\be
P_n(x)=x^n+\sum_{i=1}^{n-1}p_n^{(i)}x^i 
\ee
with $p_n^{(i)}$ coefficients, which satisfies the orthogonality 
\be
\int_{-\infty}^{\mu^2}dx\,e^{-\frac{N}{2}x^2}P_m(x)P_n(x)=h_m\delta_{mn}, 
\label{eq:OP}
\ee
and the recursion relation 
\begin{align}
xP_n(x)=P_{n+1}(x)+S_nP_n(x)+R_nP_{n-1}(x).
\label{eq:recursion}
\end{align}
Then (\ref{eq:Z(1,0)_Nicolai}) and (\ref{eq:correlator_Nicolai}) are expressed as 
\begin{align}
Z_{(1,0)}=\tilde C_NN!\prod_{n=0}^{N-1}h_n, 
\end{align}
and 
\begin{align}
\vev{\frac{1}{N}\tr\phi^{2k+1}}^{(1,0)}
=&\frac1N\sum_{n=0}^{N-1}\frac{1}{h_n}
\int_{-\infty}^{\mu^2} dx\,(\mu^2-x)^{k+\frac12}P_n(x)^2 e^{-\frac{N}{2}x^2}.
\label{eq:1ptbyOP}
\end{align}
Likewise eigenvalue distribution of the Gaussian matrix model 
\begin{align}
\rho_M^{(\text{G'})}(x)\equiv\vev{\frac1N\tr\delta(x-M)}^{(\text{G'})}
=\vev{\frac1N\sum_{i=1}^N\delta(x-x_i)}^{(\text{G'})} 
\end{align}
becomes 
\begin{align}
\rho_M^{(\text{G'})}(x)=\frac1N\sum_{n=0}^{N-1}\frac{1}{h_n}P_n(x)^2
e^{-\frac{N}{2}x^2}.
\label{eq:rhobyOP}
\end{align}
From \eqref{eq:1ptbyOP} and \eqref{eq:rhobyOP}, we find a formula of 
the one-point function as an integral of the eigenvalue distribution 
\begin{align}
\vev{\frac{1}{N}\tr\phi^{2k+1}}^{(1,0)}
=\int_{-\infty}^{\mu^2} dx\,(\mu^2-x)^{k+\frac12}\rho_M^{(\text{G'})}(x).
\label{eq:1ptbyrho}
\end{align}

In \cite{Endres:2013sda}, the orthogonal polynomials 
$P_n(x)$ are expressed in terms of the orthogonal polynomials $P^{(\text{H})}_n(x)$ in the 
standard Gaussian matrix model (without the upper bound for eigenvalues). 
$P^{(\text{H})}_n(x)$ is also a monic polynomial of degree $n$ satisfying 
\begin{align}
&\int_{-\infty}^{\infty}dx\,e^{-\frac{N}{2}x^2}P_m^{(\text{H})}(x)P_n^{(\text{H})}(x)
=h_n^{(\text{H})}\delta_{mn},\nn \\
&xP_n^{(\text{H})}(x)
=P_{n+1}^{(\text{H})}(x)+S_n^{(\text{H})}P_n^{(\text{H})}(x)
+R_n^{(\text{H})}P_{n-1}^{(\text{H})}(x) \qquad 
(S_n^{(\text{H})}=0). 
\label{eq:standardOP}
\end{align}
We determine differences 
$\tilde P_n(x)=P_n(x)-P_n^{(\text{H})}(x)$, $\tilde h_n=h_n-h_n^{(\text{H})}$, 
$\tilde S_n=S_n-S_n^{(\text{H})}=S_n$, $\tilde R_n=R_n-R_n^{(\text{H})}$ 
by taking account of the boundary effect of \eqref{eq:OP} in an iterative manner:
\begin{align}
&\tilde P_n(x) =\tilde P_n^{(1)}(x) +\tilde P_n^{(2)}(x) +\cdots, \nn \\
&\tilde S_n=\tilde S_n^{(1)}+\tilde S_n^{(2)}+\cdots, \nn \\
&\tilde R_n=\tilde R_n^{(1)}+\tilde R_n^{(2)}+\cdots, \nn \\
&\tilde h_n=\tilde h_n^{(1)}+\tilde h_n^{(2)}+\cdots, 
\end{align}
where the numbers (1), (2), $\cdots$ denote the steps of the iteration 
and turn out to correspond to the instanton numbers of the contributions~\cite{Endres:2013sda}. 
Namely, quantities with the number $(p)$ are suppressed by the factor 
$\exp\left(-4ps^{3/2}/3\right)$ as $s$ grows as in \eqref{eq:Z_pinst_dsl}. 

Applying this expansion to the eigenvalue distribution \eqref{eq:rhobyOP} with 
\begin{align}
\tilde L_n(x)\equiv\frac{\tilde P_n(x)}{P_n^{(\text{H})}(x)}
=\tilde L_n^{(1)}(x)+\tilde L_n^{(2)}(x)+\cdots,
\end{align}
we obtain 
\begin{align}
\tilde \rho_M(x) \equiv & \rho_M^{(\text{G'})}(x)-\rho_M^{(\text{G})}(x) \nn \\
=&\frac1N\sum_{n=0}^{N-1}
\frac{1}{h_n^{(\text{H})}}P_n^{(\text{H})}(x)^2e^{-\frac{N}{2}x^2}
\left(2\tilde L^{(1)}_n(x)-\frac{\tilde h_n^{(1)}}{h_n^{(\text{H})}}+\cdots\right),
\end{align}
where $\rho_M^{(\text{G})}(x)$ is the eigenvalue distribution of the standard Gaussian matrix model, 
and the ellipsis represents contribution from higher instanton numbers 
($p\geq 2$)~\cite{Endres:2013sda}.
Then the one-point function \eqref{eq:1ptbyrho} is decomposed as 
\begin{align}
\vev{\frac{1}{N}\tr\phi^{2k+1}}^{(1,0)}
=\int_{-\infty}^{\mu^2} dx\,(\mu^2-x)^{k+\frac12}\rho_M^{(\text{G})}(x)
+\int_{-\infty}^{\mu^2} dx\,(\mu^2-x)^{k+\frac12}\tilde \rho_M(x). 
\end{align}
From straightforward calculation similar to what is done in section 5 in \cite{Endres:2013sda}, 
the second term of the r.h.s. turns out to be a quantity with higher 
instanton numbers ($p\geq 2$), and can be neglected  
as far as cancellation between the zero- and  one-instanton sectors is concerned: 
\begin{align}
\left.\vev{\frac{1}{N}\tr\phi^{2k+1}}^{(1,0)}\right|_{0-{\text{inst.}}+1-{\text{inst.}}}
=\int_{-\infty}^{\mu^2} dx\,(\mu^2-x)^{k+\frac12}\rho_M^{(\text{G})}(x).
\label{eq:1ptbyrhoinGMM}
\end{align} 

By the same reason, we can replace the resolvent in 
(\ref{eq:1ptin0instbyGMM}) by that of the standard Gaussian matrix model (with the superscript (G)):
\begin{align}
\left.\vev{\frac{1}{N}\tr\phi^{2k+1}}^{(1,0)}\right|_{\text{0-inst.}}
=-\oint_C \frac{dz}{2\pi i}\,2z^{2k+2}\vev{R_M(\mu^2-z^2)}^{(\text{G})}. 
\label{eq:1ptin0-inst} 
\end{align}

%%%%%%%%%%%%%%%%%%%%%%%%%%%%%%%%%%%%%%%%%%%%%%%%%%%%%%%%%%%%%%%%%%%%%%%%%%%%%
\section{Ambiguities in the zero-instanton sector}
\label{sec:zero-inst}
\setcounter{equation}{0}
%%%%%%%%%%%%%%%%%%%%%%%%%%%%%%%%%%%%%%%%%%%%%%%%%%%%%%%%%%%%%%%%%%%%%%%%%%%%%
In \cite{Kuroki:2016ucm}, 
the all-order result of genus expansion of the one-point functions $\vev{\frac{1}{N}\tr \phi^{2k+1}}^{(1,0)}$ 
is obtained at the zero-instanton sector in the double scaling limit (\ref{eq:dsl}). 
In this section, we apply the Borel resummation technique to the result and find that ambiguities arise.  

%%%%%%%%%%%%%%%%%%%%%%%%%%%%%%%%%%%%%%%%%%%%%%%%%%%%%%%%%%%%%%%%%%%%%%
\subsection{Genus expansion to all orders}
For the resolvent of the standard Gaussian matrix model $\vev{R_M(z)}^{(\text{G})}$, 
the expression of genus expansion is obtained at arbitrary genus in the literature e.g. \cite{HT_2012}. 
In~\cite{Kuroki:2016ucm}, utilizing the result there to (\ref{eq:1ptin0-inst}), 
we have arrived at the expression
\begin{align}
&\left.\vev{\frac1N\tr\phi^{2k+1}}^{(1,0)}\right|_{\text{0-inst., univ.}} \nn \\
&=N^{-\frac23(k+2)}\frac{\Gamma\left(k+\frac32\right)}{2\pi^{\frac32}} \,s^{k+2}
\Biggl\{\sum_{h=0}^{\left[\frac13(k+2)\right]}\frac{1}{h!}\left(-\frac{1}{12}\right)^h
\frac{1}{\Gamma\left(k+3-3h\right)}s^{-3h}\ln s \nn \\
&\phantom{N^{-\frac23(k+2)}} \hspace{10mm} 
+(-1)^{k+1}\sum_{h=\left[\frac13(k+2)\right]+1}^{\infty}\frac{1}{h!}\left(\frac{1}{12}\right)^h
\Gamma\left(3h-k-2\right)s^{-3h}\Biggr\}
\label{eq:1ptgenusexp}
\end{align}
in the double scaling limit~(\ref{eq:dsl}), 
where $[x]$ denotes the greatest integer less than or equal to $x$. 
We can see that the infinite series in the bracket on the r.h.s. gives the genus expansion where the power of $g_s^2\propto s^{-3}$ counts the number of handles. 
The suffix ``univ.'' on the l.h.s. means that the most dominant nonanalytic term 
at $s=0$ is taken in the limit~(\ref{eq:dsl}) (the universal part). 
The overall factor $N^{-\frac23(k+2)}$ can be absorbed 
in the ``wave function renormalization'' of the operator $\frac1N\tr\phi^{2k+1}$. 

%%%%%%%%%%%%%%%%%%%%%%%%%%%%%%%%%%%%%%%%%%
\subsection{Borel resummation}
The second line on the r.h.s. in \eqref{eq:1ptgenusexp} is a series exhibiting factorial growth as $\frac{\Gamma(3h-k-2)}{h!}\sim (2h)!$, which is a characteristic feature 
of string perturbation series and gives further support that the matrix model describes 
a string theory in the double scaling limit \cite{Shenker:1990uf}. 

\begin{figure}[h]
\centering
\includegraphics[width=10cm, bb=0 0 800 800, clip]{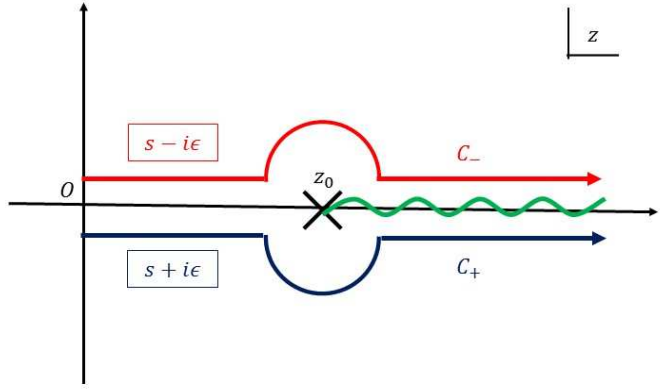}
\vspace{-4cm}
\caption{\small Integration contours $C_+$ and $C_-$ on the Borel plane.}
\label{fig:twocontours2}
\end{figure}

The factorial growth means that \eqref{eq:1ptgenusexp} is a divergent series with convergence radius zero.  
In order to try to make the series well-defined, let us apply the Borel resummation technique to \eqref{eq:1ptgenusexp}. 
It amounts to inserting  
\begin{align}
1=\frac{1}{\Gamma\left(2h+1\right)}\int_0^\infty dz\,z^{2h}e^{-z}
\label{eq:resummationidentity}
\end{align}
into \eqref{eq:1ptgenusexp} and interchanging the order of the sum on $h$ and the integral on $z$. 
Use of Stirling's formula $\Gamma(x)\sim \sqrt{2\pi}\,x^{x-\frac12}\,e^{-x}\,\left[1+\frac{1}{12x} +\cO(x^{-2})\right]$ ($x\to \infty$) leads to 
\bea
\frac{\Gamma(3h-k-2)}{h!\Gamma(2h+1)} & \sim & \frac{1}{2\sqrt{\pi}\,3^{k+\frac{5}{2}}}\left(\frac{27}{4}\right)^h h^{-k-\frac72} \nn \\
& & \times \left[1+\left\{(k+2)(k+3)-\frac{7}{12}\right\}\frac{1}{6h} +\cO(h^{-2})\right] 
\label{coeff_1pt}
\eea
for large $h$. 
In addition, a binomial coefficient $\binomi{\alpha}{h}$ with $\alpha\notin \bm Z$ asymptotically behaves 
\bea
\binomi{\alpha}{h} & = & (-1)^{h+1}\frac{\sin(\pi\alpha)}{\pi}\Gamma(\alpha+1)\, \frac{\Gamma(h-\alpha)}{h!}  \nn \\
& \sim &   (-1)^{h+1}\frac{\sin(\pi\alpha)}{\pi}\Gamma(\alpha+1)\, h^{-\alpha-1}\,\left[1+\frac{\alpha(\alpha+1)}{2h} + \cO(h^{-2})\right].
\label{binomial}
\eea
Combining these two, we can express (\ref{coeff_1pt}) as an expansion 
by binomial coefficients~\footnote
{For $m=0,1,2,\cdots$, $\binomi{k+\frac52+m}{h}$ is a quantity of $\cO\left(h^{-k-\frac72-m}\right)$ from eq.\eqref{binomial}.}:
\bea
\frac{\Gamma(3h-k-2)}{h!\,\Gamma(2h+1)} & \sim & \frac{(-1)^{h+k+1}\sqrt{\pi}}{2\cdot 3^{k+\frac52}}\,\left(\frac{27}{4}\right)^h 
\left[\frac{1}{\Gamma(k+\frac72)} \binomi{k+\frac52}{h}  \right. \nn \\
& & \hspace{5mm} \left. +\frac{12k^2+78k+125}{36}\,\frac{1}{\Gamma(k+\frac92)} \binomi{k+\frac72}{h} 
+\cO(h^{-k-\frac{11}{2}})\right]. 
\eea
Then the Borel resummed series of (\ref{eq:1ptgenusexp}) becomes 
\begin{align}
&N^{\frac23(k+2)}\left.\vev{\frac1N\tr\phi^{2k+1}}^{(1,0)}\right|_{\text{0-inst., univ.}}^{\text{ B.R.}}
\nn \\
&
=\frac{1}{4\pi}
\frac{s^{k+2}}{3^{k+\frac52}\left(k+\frac32\right)\left(k+\frac52\right)}\left[ \int_0^{\infty}dz \,\left(1-\frac{z^2}{z_0^2}\right)^{k+\frac52}e^{-z} \right. \nn \\
& \hspace{14mm} \left. + \frac{12k^2+78k+125}{36 \left(k+\frac72\right)} \int_0^{\infty}dz \,\left(1-\frac{z^2}{z_0^2}\right)^{k+\frac72}e^{-z}+\cdots\right],
\label{eq:resummation}
\end{align}
with 
\begin{align}
z_0\equiv\frac43s^{\frac32}.
\end{align}
Here the ellipsis stands for integrals containing higher powers of $\left(1-\frac{z^2}{z_0^2}\right)$ 
and lower genus contributions up to $h=\left[\frac13(k+2)\right]$. 
The former contributes to a series of ambiguities as well as the first two terms, 
whereas the latter is a finite sum providing nothing ambiguous. 
The integrals in \eqref{eq:resummation} yield ambiguities due to the cut of the integrands 
$[z_0, +\infty)$ on the integration contour. 
As depicted in Fig.~\ref{fig:twocontours2}, there are two ways to avoid the cut in the integrals of $z$, 
but the result will change depending on which choice we take. 
This means that the divergent series cannot be made well-defined (non-Borel summable) 
for the zero-instanton sector alone. 
However, by taking account of contributions from nonzero instanton numbers, there is a possibility that ambiguity arising there cancels the ambiguity from the zero-instanton sector. 
Therefore, the theory is free from ambiguity and well-defined as a whole. 
This is the idea of resurgence. 
In the following, we will actually see that a series of ambiguities from (\ref{eq:resummation}) 
cancels that from one-instanton contribution (up to the next-to-leading order). 
 
%%%%%%%%%%%%%%%%%%%%%%%%%%%%%%%%%%%%%%%%%%%%%%%%%%%%%%%%%%%%%%%%%%%%%%%%

In order to identify the precise form of ambiguities, let us give a tiny imaginary part to $s$: 
$s\rightarrow s\pm i\epsilon$ 
with $\epsilon>0$. 
For $s+i\epsilon$ ($s-i\epsilon$), the integration contour of $z$ 
in \eqref{eq:resummation} shifts slightly below (above) the positive real axis 
to $C_+$ ($C_-$) as in Fig. \ref{fig:twocontours2}. 
Then the ambiguities are given as difference between contribution from $s+i\epsilon$ ($C_+$) 
and that from $s-i\epsilon$ ($C_-$), i.e., integrated discontinuity of the integrands across the cut:
\begin{align}
&(\text{Ambiguities of (\ref{eq:resummation})}) \equiv 
\left.(\ref{eq:resummation})\right|_{s\to s+i\epsilon} - \left.(\ref{eq:resummation})\right|_{s\to s-i\epsilon} \nn \\
&=i(-1)^k\frac{1}{2\pi}
\frac{s^{k+2}}{3^{k+\frac52}\left(k+\frac32\right)\left(k+\frac52\right)}\left[
\int_{z_0}^{\infty}dz\,\left(\frac{z^2}{z_0^2}-1\right)^{k+\frac52}e^{-z} \right. \nn \\
&  \hspace{27mm} 
\left. - \frac{12k^2+78k+125}{36 \left(k+\frac72\right)} \int_{z_0}^{\infty}dz \,\left(\frac{z^2}{z_0^2}-1\right)^{k+\frac72}e^{-z}+\cdots\right] \nn \\
&=i(-1)^k\frac{1}{3^{\frac12}\pi^{\frac32}}
\frac{\Gamma\left(k+\frac32\right)}{2^{k+2}s^{\frac12k+1}}
\left[K_{k+3}(z_0) -\frac{12k^2+78k+125}{18z_0}\,K_{k+4}(z_0) +\cdots\right],
\end{align}
where the ellipsis in the last line stands for terms of modified Bessel function 
$K_{k+3+\nu}(z_0)$ with $\nu\geq 2$. 
They are accompanied with $z_0^{-\nu}$ as in the second term 
and hence suppressed by $s^{-3\nu/2}$ compared to the first term. 
By using the asymptotic form 
\begin{align}
K_{\nu}(z)\sim \sqrt{\frac{\pi}{2z}}e^{-z}\left[1+ \frac{4\nu^2-1}{8z}+{\cal O}(z^{-2})\right]\qquad 
(z\rightarrow+\infty),
\end{align}
we finally find 
\begin{align}
(\text{Ambiguities of (\ref{eq:resummation})}) 
=i(-1)^k\frac{\Gamma\left(k+\frac32\right)}{2^{k+\frac72}\pi}\,\frac{e^{-\frac43s^{\frac32}}}
{s^{\frac{k}{2}+\frac74}} \left[1-\frac{1}{8s^{\frac32}}\left(k^2+8k+\frac{185}{12}\right)
+{\cal O}(s^{-3})\right].
\label{eq:amb.in0-inst.}
\end{align}
Notice that the exponential factor $e^{-\frac43s^{\frac32}}$ is characteristic 
of the one-instanton contribution and its exponent comes from the value of the branch point $z_0$. 
This opens a profound connection between perturbative ambiguities 
and nonperturbative effects~\cite{zinnjustin}. 
In addition, ambiguities from higher powers of $\left(\frac{z^2}{z_0^2}-1\right)$ 
seem to be related to 
higher corrections by holes and handles which are created by D-branes and closed strings, respectively.  

%%%%%%%%%%%%%%%%%%%%%%%%%%%%%%%%%%%%%%%%%%%%%%%%%%%%%%%%%%%%%%%%%%%%%%%%%%%%%
\section{Ambiguities in the one-instanton sector}
\label{sec:one-inst}
\setcounter{equation}{0}
%%%%%%%%%%%%%%%%%%%%%%%%%%%%%%%%%%%%%%%%%%%%%%%%%%%%%%%%%%%%%%%%%%%%%%%%%%%%%
In the previous section, we have seen that the one-point functions of the odd-power operators provide divergent string perturbation series in the zero-instanton sector, 
and explicitly computed a series of ambiguities of its Borel resumed series 
at the leading and next-to-leading orders. 
Here we find that another series of ambiguities appears in the one-instanton sector, 
and that these two series with different origins indeed cancel each other. 

%%%%%%%%%%%%%%%%%%%%%%%%%%%%%%%%%%%%%%%%
\subsection{One-point functions in the one-instanton sector}
By using the fact that the eigenvalue distribution of the standard Gaussian matrix model becomes the (diagonal) Airy kernel in the double scaling limit (\ref{eq:dsl})~\cite{Nishigaki:2014ija,TW}: 
\begin{align}
N^{\frac13}\rho_M^{(\text{G})}(x) \to
K_{\text{Ai}}(\xi,\xi)\equiv \text{Ai}'(\xi)^2-\xi\text{Ai}(\xi)^2 
\label{eq:Airykernel}
\end{align}
with $x=2+N^{-\frac23}\xi$, we obtain the expression of (\ref{eq:1ptbyrhoinGMM}) in the double scaling limit as 
\begin{align}
N^{\frac23(k+2)}\left.\vev{\frac{1}{N}\tr\phi^{2k+1}}^{(1,0)}
\right|_{0-{\text{inst.}}+1-{\text{inst.}}}
\rightarrow\int_{-\infty}^{s} d\xi\,(s-\xi)^{k+\frac12}K_{\text{Ai}}(\xi,\xi).
\end{align}
From the relation of the Nicolai mapping $2+N^{-\frac23}\xi=x=\mu^2-\lambda^2$, we see that the region $[a,b]$ for $\lambda$ in \eqref{eq:instantonsectors} is mapped to $(-\infty,0]$ for $\xi$,  
while the region $[0,a)$ is mapped to $(0,s]$ 
in the double scaling limit. 
($\xi=s$ corresponds to the location of the instanton $\lambda=0$.) 
Thus the latter gives contribution from the one-instanton sector: 
\begin{align}
N^{\frac23(k+2)}\left.\vev{\frac{1}{N}\tr\phi^{2k+1}}^{(1,0)}\right|_{1-{\text{inst.}}}
\to \int_{0}^{s} d\xi\,(s-\xi)^{k+\frac12}K_{\text{Ai}}(\xi,\xi).
\label{eq:one-ptinone-inst}
\end{align} 

%%%%%%%%%%%%%%%%%%%%%%%%%%%%%%%%%%%%%%%%
\subsection{Saddle point method}
We consider contribution to the integral (\ref{eq:one-ptinone-inst}) from $\xi\sim s\gg1$, 
which is expected to be comparable with the ambiguities (\ref{eq:amb.in0-inst.}). 
From asymptotic behavior of the Airy function, we find that the Airy kernel behaves as 
\be
K_{\text{Ai}}(\xi,\xi)\sim \frac{e^{-\frac43\xi^{\frac32}}}{8\pi\xi}\,
\left[1 -\frac{17}{24\,\xi^{\frac32}} +{\cal O}\left(\xi^{-3}\right)\right] \qquad (\xi\to \infty),
\label{airykernel}
\ee
where we take account of the expansion up to the next-to-leading order 
in order to compare the result in the zero-instanton sector (\ref{eq:amb.in0-inst.}). 
This leads us to consider the following integral 
\be
\int_0^sd\xi\,(s-\xi)^{k+\frac12}K_{\text{Ai}}(\xi,\xi) 
\quad \rightarrow \quad \frac{1}{8\pi}\int^sd\xi\,e^{-f(\xi)},
\label{eq:fdef}
\ee
with 
\begin{align}
f(\xi)\equiv \frac43\xi^{\frac32}-\left(k+\frac12\right)\ln(s-\xi)+\ln\xi 
+\frac{17}{24}\xi^{-\frac32}+{\cal O}(\xi^{-3}).
\label{eq:exponent}
\end{align}
For a while, we do not specify the lower bound of the integral region in (\ref{eq:fdef}), 
since our concern is contribution around $\xi=s$. 
Let us evaluate the integral (\ref{eq:fdef}) by a saddle point method.   
The saddle point around $\xi=s$ is found to be   
\begin{align}
\xi_*=s+\frac{k+\frac12}{2}s^{-\frac12}v(s), \qquad 
v(s)\equiv 1-\frac{k+\frac52}{4}s^{-\frac32} +\cO(s^{-3}),
\label{eq:xisaddlept}
\end{align}
where $v(s)$ represents corrections by subleading contributions. 
We see that the saddle point corresponding to the instanton $\xi_*=s$ slightly shifts 
due to the presence of the operator depending on $k$. 

In the standard saddle point method, we first find a saddle point and rotate 
an original integration contour so that it will go along the steepest 
descent path passing through the saddle point. 
In our case, the original integration contour is not an infinite line, but a finite interval $[0,s]$,  
which is major difference from the standard case.   
The region $[0,s]$ is along the steepest descent path, 
but it terminates at the branch point $\xi=s$.  
The saddle point \eqref{eq:xisaddlept} is not in the interval $[0,s]$, but on the branch cut $[s,\infty)$. 
This can be treated by the shift $s\to s \pm i\epsilon$ as in section~\ref{sec:zero-inst}. 
Here we rotate the contour by the angle $\pi$ (or $-\pi$) around $\xi=s$. 
The rotated contour goes in the opposite direction decreasing $\xi$ 
on the real axis, passes through $\xi_*$ and ends at $\xi=s$. 
We should choose the $\pi$ rotation or the $-\pi$ rotation of the contour of $\xi$ 
in accordance with $s\rightarrow s+ i\epsilon$ or $s\rightarrow s- i\epsilon$, respectively. 

Here it is worth noticing that 
what resurgence implies in the present setting. We first note that 
\eqref{eq:1ptbyrhoinGMM} itself is a well-defined real quantity without ambiguity, and  
we are interested in it after taking the double scaling limit. 
Because of technical difficulty, without explicit computation of the integral \eqref{eq:1ptbyrhoinGMM}, 
we are trying to deduce its expression 
in the double scaling limit in the form of trans-series. 
Quantity in each instanton sector would have 
ambiguities, but resurgence ensures that reflecting the well-definedness 
of the original expression, such ambiguities are expected to cancel among instanton sectors. 
Even if the integration region does not contain a saddle point, 
we should develop perturbation theory around it in order to 
construct trans-series. As we will see later, information of the original contour 
can be included in how to rotate the contour and the end point of the rotated contour. 

Let us go back to the computation. 
In calculating $f(\xi_*)$, we should use the shift $s\to s\pm i\epsilon$ to the term of $\ln(s-\xi_*)$: 
\begin{align}
\ln(s-\xi_*) \rightarrow \ln(s\pm i\epsilon-\xi_*)=
\ln(\xi_*-s)\pm i\pi \qquad (s\rightarrow s\pm i\epsilon). 
\label{eq:origin}
\end{align}
Then 
\begin{align}
f(\xi_*) =& \pm i\pi\left(k+\frac12\right) +\tilde{f}(\xi_*), \\
\tilde f(\xi_*)\equiv & \frac43 s^{\frac32}
+\left(\frac{k}{2}+\frac54\right)\ln s
-\left(k+\frac12\right)\ln\left(\frac{k+\frac12}{2}\right)+k+\frac12  \nn \\
 & +\frac18\left\{\left(k+\frac12\right)\left(k+\frac92\right)+\frac{17}{3}\right\}s^{-\frac32} 
+{\cal O}\left(s^{-3}\right), 
\label{eq:saddleptvalue}
\end{align}
while $\xi$-derivatives have no ambiguity: 
\begin{align}
f''(\xi_*)=& \frac{4s}{k+\frac12}\frac{1}{v(s)^2} +s^{-\frac12} +\cO(s^{-2}), \nn \\
f^{(n)}(\xi_*)= &(-1)^n \left(k+\frac12\right) \Gamma(n)
\left(\frac{2s^{\frac12}}{k+\frac12}\frac{1}{v(s)}\right)^n
\left(1+\cO(s^{-3})\right) \qquad (n\geq 3).
\label{eq:fderivatives}
\end{align}
The Taylor expansion of $f(\xi)$ is summed up as 
\begin{align}
f(\xi) = & f(\xi_*) +\sum_{n=2}^\infty \frac{1}{n!} f^{(n)}(\xi_*)x^n \nn \\
= & f(\xi_*) +\frac{2s^{\frac12} x}{v(s)} -\left(k+\frac12\right)\ln\left(1+\frac{2s^{\frac12}}{k+\frac12} \frac{x}{v(s)} \right) 
+\frac12s^{-\frac12} x^2 +\cO(s^{-3})
\label{fxi_expansion}
\end{align}
with $x=\xi-\xi_*$. 
Looking at the factor of the Gaussian integral $f''(\xi_*)$ in (\ref{eq:fderivatives}), 
we can regard $x$ as a quantity at most $\cO(s^{-\frac12})$. From this, 
we find that all the terms in the Taylor expansion are of the same order and should be kept.  

By changing the integration variable to 
\be
t=\frac{2s^{\frac12}}{v(s)}x
\ee 
in order to zoom in the vicinity of the saddle point, 
the upper bound of the integral $x=s-\xi_*$ still remains a finite value $t=-k-\frac12$, 
whereas the lower bound becomes far away from the saddle point by $\cO(s^{\frac12})$.  
The integral we should evaluate becomes~\footnote{
We can check that the integrand does not depend on the contour rotation by $\pi$ or $-\pi$. 
Setting 
$s-\xi=re^{i\theta}$, $\theta$ is supposed to rotate from 0 to $\pm \pi$ in accordance with 
$s\to s\pm i\epsilon$ as mentioned before. 
Then $t$ becomes 
$t=-k-\frac12-\frac{2s^{\frac12}}{v(s)}re^{i\theta}$. 
Only the subtle factor in the integrand $\left(1+\frac{t}{k+\frac12}\right)^{k+\frac12}$ is written as 
$\left(e^{\mp i\pi}
\frac{2s^{\frac12}}{\left(k+\frac12\right)v(s)}re^{i\theta}\right)^{k+\frac12}$ 
due to $-s\mp i\epsilon=e^{\mp i\pi}\,s$.
Hence it becomes the same irrespective of rotating $\theta$ by $\pi$ or $-\pi$.}   
\begin{align}
&\frac{1}{8\pi}\int^sd\xi\,e^{-f(\xi)} \nn \\ 
&= \pm \frac{i}{8\pi}(-1)^ke^{-\tilde f(\xi_*)}\frac{v(s)}{2s^{\frac12}}
\int_{\infty}^{-k-\frac12} dt\,\left(1+\frac{t}{k+\frac12}\right)^{k+\frac12}e^{-t}
\left[1-\frac{v(s)^2}{8s^{\frac32}}t^2 +{\cal O}\left(s^{-3}\right)\right] \nn \\
&= \mp i(-1)^k
\frac{\Gamma\left(k+\frac32\right)}{2^{k+\frac92}\pi}\,
\frac{e^{-\frac43s^{\frac32}}}{s^{\frac{k}{2}+\frac74}}
\left[1- \frac18\left(k^2+8k+\frac{185}{12}\right)s^{-\frac32} +{\cal O}(s^{-2})\right].
\label{1-inst_result}
\end{align}
We end up with
\begin{align}
&(\text{Ambiguities of (\ref{eq:one-ptinone-inst})}) 
\equiv \left.(\ref{1-inst_result})\right|_{s\rightarrow s+i\epsilon}
-\left.(\ref{1-inst_result})\right|_{s\rightarrow s-i\epsilon} \nn \\
&= -i(-1)^k
\frac{\Gamma\left(k+\frac32\right)}{2^{k+\frac72}\pi}\,\frac{e^{-\frac43s^{\frac32}}}{s^{\frac{k}{2}+\frac74}}
\left[1- \frac18\left(k^2+8k+\frac{185}{12}\right)s^{-\frac32}+{\cal O}(s^{-3})\right],
\label{eq:amb.in1-inst.}
\end{align}
which precisely cancels the series of ambiguities in the zero-instanton sector \eqref{eq:amb.in0-inst.} 
regarding the leading and next-to-leading contributions. 
In the above derivation, we identify the origin of ambiguities as the saddle point value 
of the integrand, more precisely, of the operator $\frac{1}{N}\tr\phi^{2k+1}$. 
In fact, 
the imaginary ambiguities come from the logarithmic term (\ref{eq:origin}) 
whose origin is the operator $(s-\xi)^{k+\frac12}$ after the Nicolai mapping. 
It is reasonable because the partition function or even-power operators $\frac1N\tr\phi^{2k}$ 
do not have any ambiguity and hence the existence of ambiguities must not depend 
on the eigenvalue distribution or the Airy kernel, but on the kind of operators. 

In (\ref{fxi_expansion}), we manage to sum up the Taylor series to obtain the logarithmic term 
which is a key to quickly derive the Gamma function $\Gamma\left(k+\frac32\right)$ 
in the ambiguities (\ref{1-inst_result}). 
If we perform the ordinary saddle point calculation, i.e., Gaussian integral over the whole real axis 
by bringing down the higher terms of $n\geq 3$ in (\ref{fxi_expansion}) from the exponent, 
the Gamma function will appear in the form of the asymptotic series as $k+\frac12$ grows~\footnote
{From (\ref{eq:fderivatives}), $x=\xi-\xi_*$ can be regarded as a quantity at most $\cO\left(\left(k+\frac12\right)^{\frac12}\right)$ in the Gaussian integral 
with respect to the $k$-dependence. 
Thus the $n$-th order term in (\ref{fxi_expansion}) is suppressed as $\cO\left(\left(k+\frac12\right)^{1-\frac{n}{2}}\right)$.}. 
A similar situation was found in \cite{Fujimori:2016ljw} 
where difference between a Gamma function factor 
and its form of Stirling's formula is supplemented by terms whose order is higher 
than quadratic. There, it is necessary to go beyond the one-loop determinant 
since  quasi zero-modes appear.  In our case, there is no such zero-mode 
since $f''(\xi_*)$ is positive definite, but still all order terms needs to be taken into account 
in order to confirm resurgence. It would be interesting that in both cases 
we need all order terms in the saddle point method for different reasons. 

In order to derive the trans-series around the instanton saddle 
for the finite interval, we took a prescription to rotate the integration contour 
so that it will pass the saddle point with taking care of 
the direction and the end point of the original contour. 
We have explicitly seen that this prescription realizes the cancellation not only at the leading order 
but also at the next-to-leading order. 
This result supports validity of our prescription. 
Some powerful technique will be necessary to check the cancellation to all orders 
or at the level of higher instanton numbers.   
It is also desirable to 
give more justification of this prescription from the viewpoint 
of general theory on resurgence applied to an interval.    

Finally, concerning our motivation mentioned in Introduction, we make a comment on a 
relation to physics, in particular spontaneous supersymmetry breaking. 
As shown in~\cite{Endres:2013sda} (eqs. (5.25) and (5.26) there~\footnote{
The variable $t$ there should be read as $s/4$.}), 
its order parameter is given by the Airy kernel (\ref{eq:Airykernel}) as
\begin{align}
N^{\frac43}\vev{\frac1N\tr(\phi^2-\mu^2)}^{(1,0)}
=K_{\text{Ai}}(s,s) + \cdots,
\label{eq:orderparam}
\end{align}
where the ellipsis stands for contribution from higher instantons. 
As explicitly seen from (\ref{airykernel}), this expression can be interpreted as contribution of the instanton, namely an isolated eigenvalue at the top of the potential. 
It would be important to understand physical aspects of a connection between 
the ambiguity computed here and the order parameter (\ref{eq:orderparam}). 

%%%%%%%%%%%%%%%%%%%%%%%%%%%%%%%%%%%%%%%%%%%%%%%%%%%%%%%%%%%%%%%%%%%%%%%%%%%%%
\section{Conclusions and discussions}
\label{sec:discussion}
%%%%%%%%%%%%%%%%%%%%%%%%%%%%%%%%%%%%%%%%%%%%%%%%%%%%%%%%%%%%%%%%%%%%%%%%%%%%%
In this paper, we have investigated resurgence structure in the one-point functions 
of nonsupersymmetric operators in the supersymmetric double-well matrix model 
which is proposed as nonperturbative formulation of two-dimensional type IIA superstring theory. 
In the zero-instanton sector, the Borel resummation technique is applied 
to a divergent string perturbation series, 
and a series of ambiguities arises depending on how to avoid the cut on the Borel plane. 
In the one-instanton sector, a special care is necessary for the integration contour 
of a finite interval which does not pass through a saddle point. 
Another series of ambiguities arise from the integrand itself evaluated at the saddle point. 
We have confirmed that these two series of ambiguities cancel each other 
both at the leading and next-to-leading order. 
Our prescription of the integration contour in the one-instanton sector 
is worth studying further and needs to be understood from the viewpoint of resurgence theory 
extended to cases of integrals over finite intervals.  

Another interesting question is how resurgence structure changes for other correlation functions 
in the same model. In fact, we have developed a derivation of multi-point functions of odd-power operators 
in~\cite{Kuroki:2016ucm} by using the result in the Gaussian matrix model~\cite{HT_2012}. 
There, it should be possible to read off large-order behavior of genus expansion for the two-point 
(or multi-point) functions 
of the odd-power operators~\cite{paperII}. 
It would be interesting to check that it again shows the stringy growth of the expansion coefficient as 
$(2h)!$ and to find structure of singularities on the Borel plane. On the other hand, in the one-instanton 
sector, we need to treat off-diagonal components of the Airy kernel $K_\text{Ai}(\xi,\eta)$, and 
it is expected that we find richer resurgence structure with variety of saddle points and 
steepest descent paths (Lefschetz thimbles). 

It is also anticipated that our analysis here can be extended to other models where 
the Nicolai mapping is available. If they are mapped to the Gaussian matrix model 
via the Nicolai mapping, correlation functions will be calculated from the result 
in the Gaussian matrix model as in \eqref{eq:1ptfullbyGMM}. 
If one is interested in the soft edge scaling limit, 
studies on the Airy kernel and resurgence structure in this paper 
will be useful there as well.

%%%%%%%%%%%%%%%%%%%%%%%%%%%%%%%%%%%%%%%%%%%%%%%%%%%%%%%%%%%%%%%%%%
\section*{Acknowledgements}
%%%%%%%%%%%%%%%%%%%%%%%%%%%%%%%%%%%%%%%%%%%%%%%%%%%%%%%%%%%%%%%%%%
We would like to thank Toshiaki Fujimori, Tatsuhiro Misumi,  Shinsuke Nishigaki, Norisuke Sakai, 
Yuya Tanizaki,  Mithat \"Unsal and Sen Zhang for useful discussions and comments. 
The work of T.~K. is supported in part by a Grant-in-Aid for Scientific Research (C), 16K05335. 
The work of F.~S. is supported in part by a Grant-in-Aid for Scientific Research (C), 25400289. 
We would like to thank the Yukawa Institute for Theoretical Physics 
at Kyoto University for hospitality during the workshop YITP-W-17-08 
"Strings and Fields 2017," and RIKEN during the workshop 
"RIMS-iTHEMS International Workshop on Resurgence Theory,"
where part of this work was carried out.

%%%%%%%%%%%%%%%%%%%%% References %%%%%%%%%%%%%%%%%%%%%%%%%%%%%%%%%%%%%%%

\end{document}